\documentclass[final,sort&compress]{mohproc}
\layoutstyle{6x9}

\newcommand{\udrag}{nN$\cdot$s$\cdot$m$^{-1}$}

\begin{document}

\title{Damping Properties of the Hair Bundle}



\author{Johannes Baumgart}{
   address={Institute of Scientific Computing, Department of Mathematics,\\ Technische Universit\"at Dresden, 01062 Dresden, Germany;\\
\emph{(current address:} Max Planck Institute for the Physics of Complex Systems,\\ N\"othnitzer Str. 38, 01187 Dresden, Germany\emph{)}}}

\author{Andrei S. Kozlov}{
  address={Howard Hughes Medical Institute and Laboratory of Sensory Neuroscience,\\ The Rockefeller University, 1230 York Avenue, New York, New York 10065, USA}}

\author{Thomas Risler$^{\ddag,}$}{
  address={Institut Curie, Centre de Recherche, F-75005 Paris, France}
  ,altaddress={UPMC Univ Paris 06, UMR 168, F-75005 Paris, France}
  ,altaddress={CNRS, UMR 168, F-75005 Paris, France}}

\author{A.~J.~Hudspeth}{
  address={Howard Hughes Medical Institute and Laboratory of Sensory Neuroscience,\\ The Rockefeller University, 1230 York Avenue, New York, New York 10065, USA}
}

\begin{abstract}

The viscous liquid surrounding a hair bundle dissipates energy and dampens oscillations, which poses a fundamental physical challenge to the high sensitivity and sharp frequency selectivity of hearing. To identify the mechanical forces at play, we constructed a detailed finite-element model of the hair bundle. Based on data from the hair bundle of the bullfrog's sacculus, this model treats the interaction of stereocilia both with the surrounding liquid and with the liquid in the narrow gaps between the individual stereocilia.

The investigation revealed that grouping stereocilia in a bundle dramatically reduces the total drag. During hair-bundle deflections, the tip links potentially induce drag by causing small but very dissipative relative motions between stereocilia; this effect is offset by the horizontal top connectors that restrain such relative movements at low frequencies. For higher frequencies the coupling liquid is sufficient to assure that the hair bundle moves as a unit with a low total drag. 

This work reveals the mechanical characteristics originating from hair-bundle morphology and shows quantitatively how a hair bundle is adapted for sensitive mechanotransduction.

\end{abstract}

\maketitle

\section{Introduction}
\begin{figure}[htp]
  \includegraphics{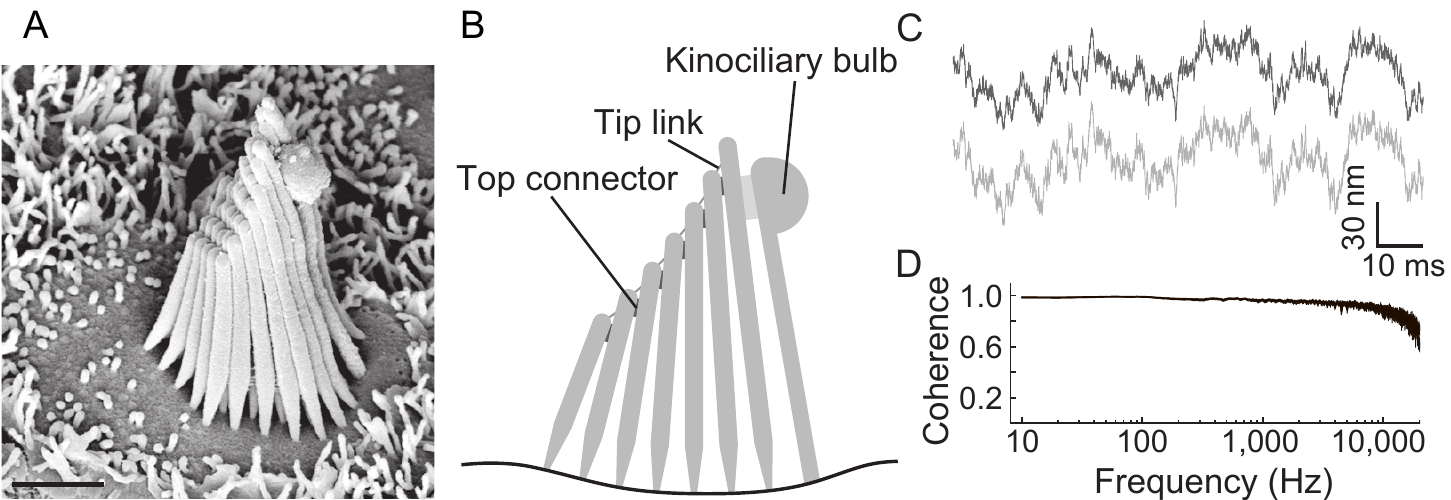}
\caption{\label{fig:ref} Structure and movement of a hair bundle. (A) a scanning
electron micrograph of a hair bundle from the bullfrog's sacculus illustrates an array
of closely apposed, cylindrical stereocilia separated by small gaps. The calibration bar corresponds to 2~\textmu{}m. (B) a schematic diagram depicts a single file of stereocilia in a hair bundle's plane of symmetry. (C) Simultaneous records of the movement at the two edges of a thermally excited hair bundle convey the impression of highly coherent motion, which is confirmed by the calculated coherence spectrum (D).}
\end{figure}

The key initial step of hearing~\cite{Hudspeth1989}, the transformation from mechanical motion into an electrical signal, takes place in the hair bundles. Each hair bundle comprises many apposed elastic stereocilia that are located in a viscous liquid that dissipates energy (Fig.~\ref{fig:ref}~A--B). In this work we focus on the mechanical environment of the mechanotransduction process.

Simultaneous measurements of the stereociliary motion at the bundle's opposite edges reveal a high coherence (Fig.~\ref{fig:ref}~C--D, \cite{Kozlov2007}). This unitary movement of the bundle is significantly less dissipative than the displacement pattern with relative motion~\cite{Kozlov2011}. We studied in detail the mechanics of the close apposition of the stereocilia to understand the forces at play, which must be sufficiently high to work against the pivotal stiffness of the individual stereocilia and to couple them. Our findings confirm previous observations by Karavitaki and Corey~\cite{Karavitaki2010} and provide futher insights into the mechanics.

In the following, we present estimates for the drag between closely apposed stereocilia and of the bundle due to the external liquid. These results are discussed in the context of a detailed finite-element model of the hair bundle. Finally, a simplified mechanical representation of the mechanics of the passive hair bundle is proposed.

\section{Drag forces in the bundle}

At the length scales of the hair bundle and at audible frequencies, the liquids of the inner ear are nearly incompressible and have low inertia. Furthermore, non-linearities can be neglected as amplitudes of motion are small compared to the geometrical dimensions. Experimental data show that even the displacement between two adjacent stereocilia is small compared to their relative separation~\cite{Kozlov2011}. This allows us to simplify the governing equations and to find analytical estimates for the drag caused by the viscous liquid.

\subsection{Relative motion between stereocilia}

The squeezing mode of motion induces the highest drag. In the gap between the cylinders the velocity profiles over the gap height are quadratic with a large velocity gradient next to the surfaces. An analytical approximation was derived for small gaps with the minimum gap distance varying linearly over the cylinders' length. For a relative motion between the two cylinders this reads
\begin{equation}\label{eq:drag:squeeze}
c_{\mathrm{squeeze}}=\pi\,\eta\,h\,\frac{\xi ^2 (3+\xi ) \chi ^3}{(1+\xi )^3}
\quad \textrm{with} \quad 
\xi = \sqrt{\frac{g_{\mathrm{t}}}{g_{\mathrm{b}}}}
\quad\textrm{and}\quad
\chi = \sqrt{\frac{r}{g_{\mathrm{t}}}}\;.
\end{equation}
Here $h$ is the cylinders' height, $r$ their radius, $g_{\mathrm{t}}$ and  $g_{\mathrm{b}}$ the wall-to-wall distance at the tip and bottom, and $\eta$ the fluid's dynamic viscosity. A similar drag coefficient estimation was conducted by Zetes~\cite{Zetes1997} for parallel cylinders. Further details are given in~\cite{Kozlov2011,Baumgart2010b}.

\begin{figure}[htp]
  \includegraphics{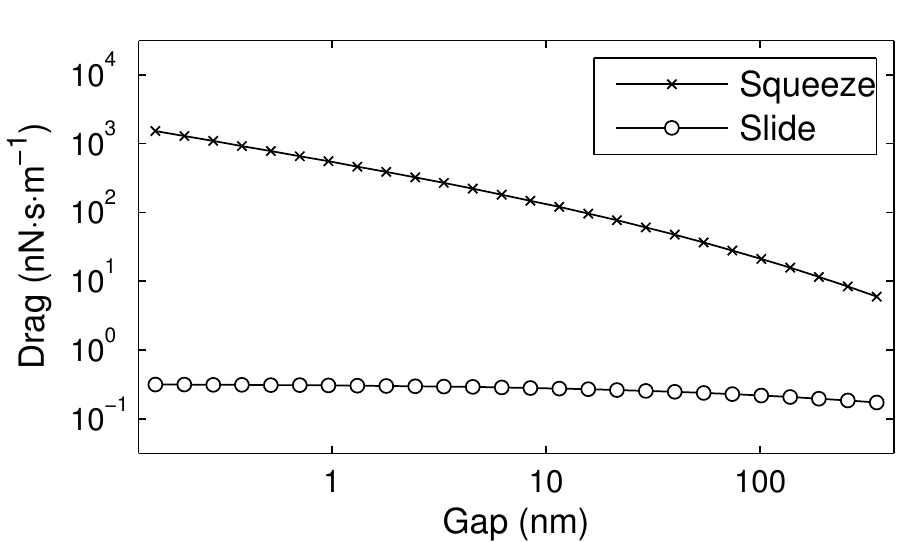}
\caption{\label{fig:cyl} Drag coefficient of pivoting cylinders moving in the plane of their axes. The cylinders' height measure $h=5$~\textmu{}m, their radius $r=0.19$~\textmu{}m, and their wall-to-wall distance at the bottom $g_{\mathrm{b}}=0.4$~\textmu{}m. The wall-to-wall distance at the tips $g_{\mathrm{t}}$ varies along the abscissa. The fluid's dynamic viscosity is set to $\eta=1$~mPa$\cdot$s. The drag of two cylinders moving toward their common center causes a high drag due to the squeezing flow in the gap (Squeeze, Eq.~\ref{eq:drag:squeeze}). If the two  cylinders are moving in the same direction, the related drag is lower by orders of magnitude (Slide, Eq.~\ref{eq:drag:slide}).}
\end{figure}

In Fig.~\ref{fig:cyl} the drag coefficient as computed from Eq.~\ref{eq:drag:squeeze} is given for a typical stereociliary geometry. The drag diverges with decreasing wall-to-wall distance at the tip and decreases by about a factor of five per decade around the typical wall-to-wall distance of 10~nm.  

\subsection{Sliding motion of stereocilia}          

For the sliding motion the drag is induced by the shear of the liquid between
the two stereocilia. If the common translatory motion is removed, the only remaining motion that induces drag is their relative motion as both cylinders move along their axis but in opposite directions. Based on the analytical solution by Hunt \emph{et al.}~\cite{Hunt1994} for the drag of a cylinder moving along its axis parallel to a plane wall and taking into account the lever-arm ratios of the pivotal motion, the drag coefficient reads
\begin{equation}\label{eq:drag:slide}
c_{\mathrm{slide}}=2\,\pi\,\eta\,\frac{r^2}{h^2}\int_{z=0}^h\frac{1}{\mathrm{arccosh}\left(g_\mathrm{o}(z)/r+1\right)}\,\mathrm{dz}
\quad \textrm{with} \quad
g_\mathrm{o}(z) = g_{\mathrm{b}} - \frac{g_{\mathrm{b}}-g_{\mathrm{t}}}{h}\,z\;.
\end{equation}
The integral was evaluated numerically for given geometries (Fig.~\ref{fig:cyl}). For this sliding motion the drag values with about 0.2~\udrag{} are significantly lower than for the relative modes  and almost independent of the wall-to-wall distance.


\section{Detailed model}
We implemented a realistic physical model of the hair bundle with an appropriate representation of the fluid-structure interactions that is able to identify the relevant physical effects~\cite{Kozlov2011,Baumgart2010b,Baumgart2009a}. To solve the boundary-value problem of the displacement fields of solid and liquid, we employed the finite-element method in three-dimensional space. In contrast to previous models~\cite{Geisler1993,Pickles1993,Zetes1997,Cotton2004a} this can resolve the liquid motion between and around the stereocilia simultaneously, without any constrain on the known geometry~\cite{Jacobs1990}, by discretizing the complex geometry with hexahedral elements (Fig.~\ref{fig:fem}).

\wrappedfigure[L]{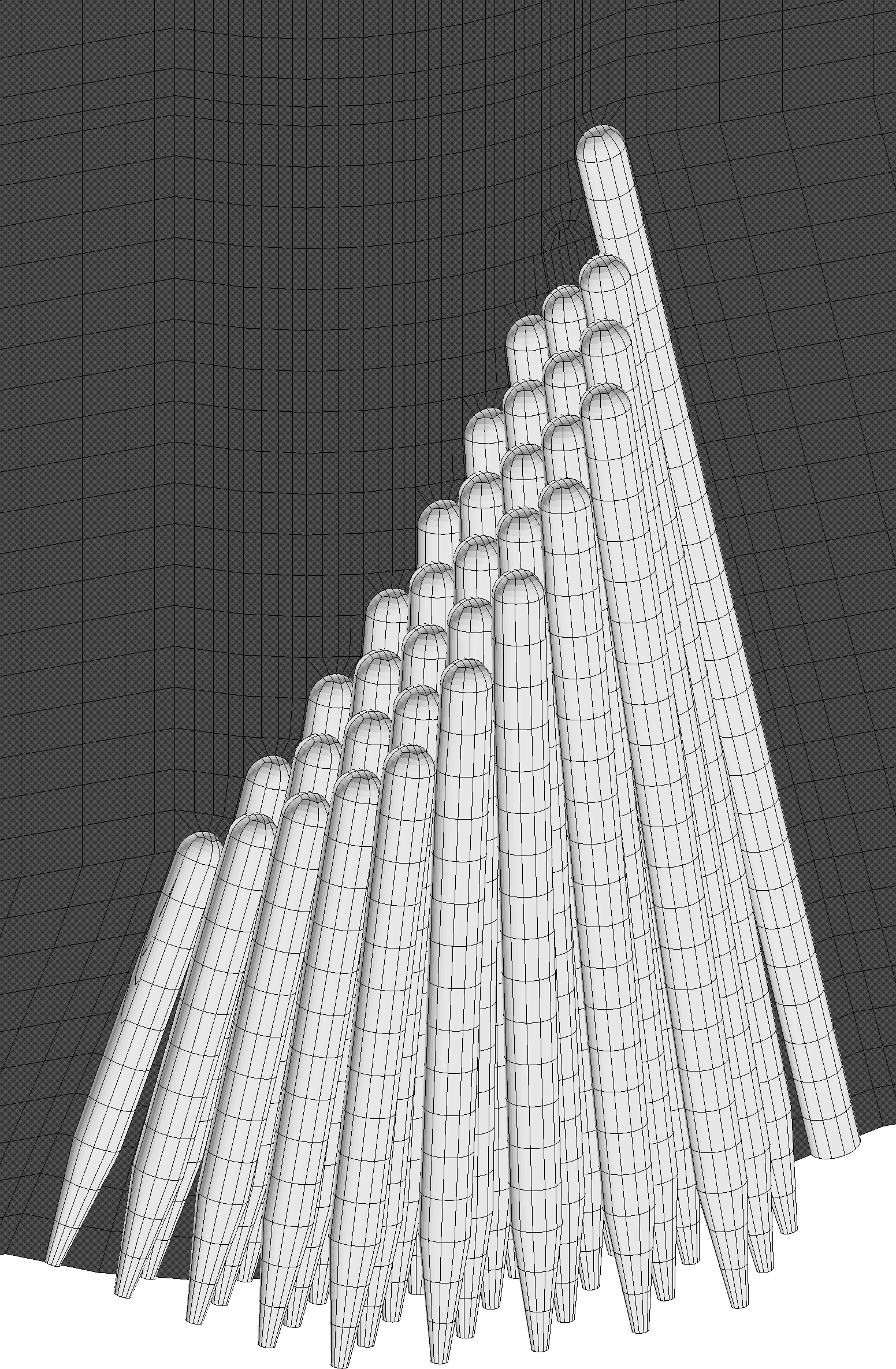}{2.585in}{3.950in}{fig:fem}
{Mesh of the finite-element hair-bundle model with the mesh of the liquid shown in the symmetry plane.}
The stiffness of the horizontal top connectors can be determined from coherence measurements of hair bundles without tip links. Furthermore, recent stiffness measurements on these bundles without tip links~\cite{Kozlov2011} yield a pivotal stiffness for the individual rootlet of only 10~aN$\cdot$m$\cdot$rad$^{-1}$. The stiffness values of the tip links and top connectors are set to 1 and 20~mN$\cdot$m$^{-1}$, respectively, to match the experimental observations.

\begin{figure}[htb]
\includegraphics{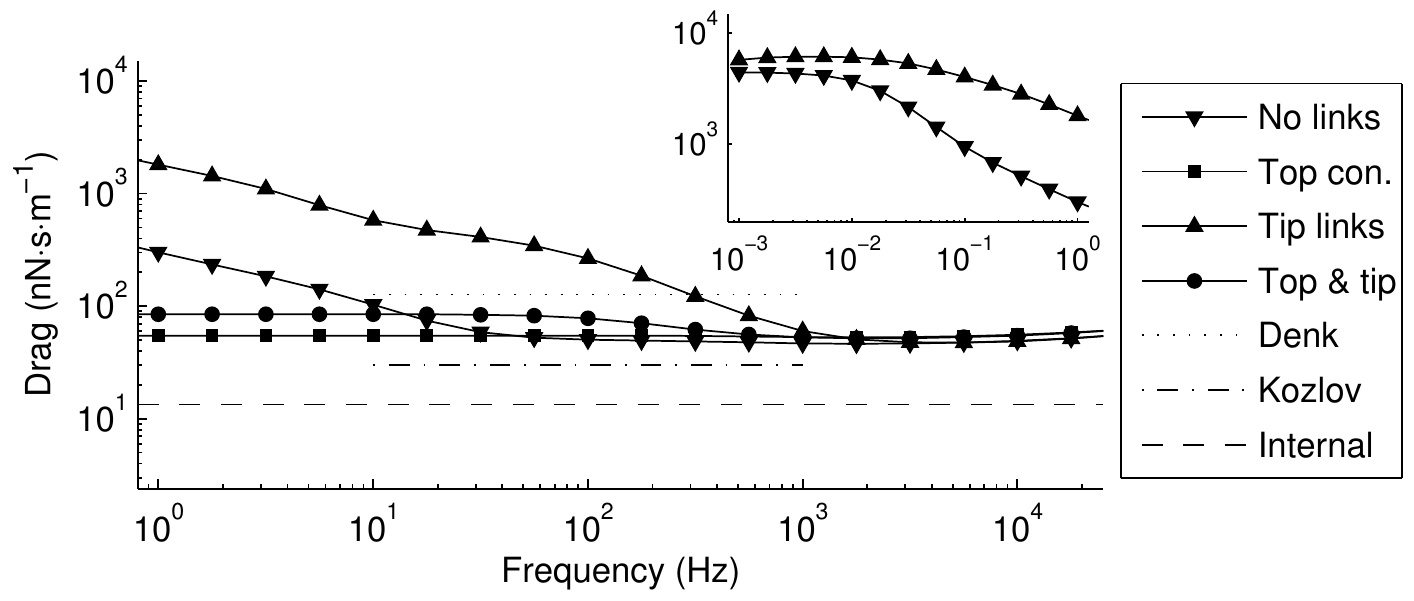}
\caption{\label{fig:drag}Drag of the hair bundle at the kinociliary bulb. Stereocilia are always coupled by the viscous liquid and elastically coupled only by horizontal top connectors (Top con.), tip links (Tip links), both types of elastic links (Top \& tip), and no elastic links (No links). As comparison are given: the minimal drag for a bundle without surrounding liquid (Internal), and experimental data from Kozlov \emph{et al.}~\cite{Kozlov2011} from hair bundles without tip links and from Denk \emph{et al.}~\cite{Denk1989} from intact hair bundles. The inset, which has axis labels identical to those of the main panel, displays the behavior at very low frequencies.}
\end{figure}

An external force is applied at the kinociliary bulb of the model in the stimulus direction to compute the drag as ratio of force divided by velocity (Fig.~\ref{fig:drag}). The purely liquid-coupled bundle has a drag coefficient of up to 4,400~\udrag{} at a frequency of 1~mHz. The drag coefficient decays as the coupling forces of the viscous liquid between the stereocilia overcome the pivotal stiffness of the stereocilia. For frequencies of 0.1~kHz and higher the drag is around 50~\udrag{}. This value is slightly higher than the experimentally measured drag of $30\pm13$~\udrag{}~\cite{Kozlov2011} of a bundle without tip links and the calculated drag of 29~\udrag{} for a hemi-ellipsoid displaced at the tip pivoting around one of the minor axes~\cite{Perrin1934}. This implies a minor contribution by internal losses. If the external liquid is removed in the model and the bundle moves coherently, the drag coefficient becomes around 13~\udrag{}. The equivalent drag of a liquid-filled hemi-ellipsoid displaced at the tip and subjected to uniform shear is 2.4~\udrag{}, which is a lower bound as it is unlikely that the bundle with internal liquid shears uniformly.

The top connectors fully block the relative modes and the associated drag. The value is constant around 55~\udrag{}. A bundle with only tip links as elastic coupling elements between stereocilia exhibits the highest damping compared to the other three cases. Below the frequency of 20~mHz the drag is around 6,000~\udrag{}. With increasing frequency the drag always exceeds that of the purely liquid-coupled bundle. The oblique tip links transfer about the same displacement amplitude from the tall stereocilium to the next shorter one, but the lever arms of the center of the pivotal motion with respect to the tip-link connection point differ. The shorter stereocilium displaces at a lower height and therefore rotates further, causing additional relative motion with associated drag. From around 3~kHz to higher frequencies the viscous coupling of the liquid overcomes the tip-link stiffness and the drag aligns with the purely liquid-coupled solution. The drag of a bundle with both types of elastic links between stereocilia follows closely the drag of a bundle with only top connectors for frequencies above 1~kHz. For frequencies from 0.1~Hz to 0.1~kHz the relative motion induced by the tip links increases the drag to 85~\udrag{}. 

\section{Conclusions}

\wrappedfigure[r]{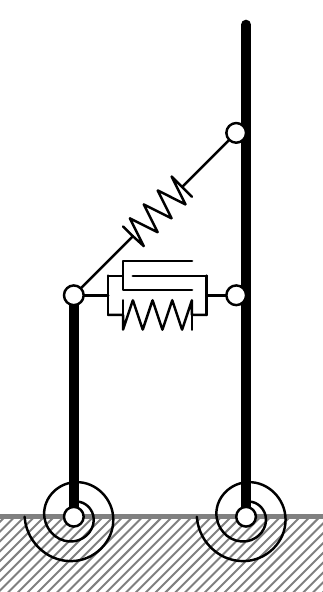}{3.25cm}{6cm}{fig:sketch}
{Sketch of bundle mechanics.}
If two closely apposed stereocilia move to their common center the related drag coefficient is larger by orders of magnitude compared to a unitary motion. A coherent and low-dissipation motion can be assured by the horizontal top connectors or -- if the frequency is sufficiently high -- by the viscous liquid filling the gap. 

Based on the detailed model the essential mechanics can be simplified as given in Fig.~\ref{fig:sketch}. The two rigid stereocilia are interconnected by an oblique tip-link stiffness, a horizontal top-connector stiffness, and a damper representing the coupling liquid. At the rootlet the  pivotal stiffness might be put in series with a pivotal damper to mimic the drag by the external liquid. The parameter values of the springs and dampers should be chosen such that the coupling forces are higher for the horizontal part than for the oblique tip links. 

It is remarkable how the characteristics of the viscous liquid are transformed into an asset in the geometrical arrangement of the hair bundle to reduce the total drag and couple the stereocilia coherently.

\begin{theacknowledgments}
We thank A. J. Hinterwirth for assistance in constructing the interferometer and B. Fabella for programming the experimental software; C. P. C. Versteegh for drag measurements on a scaled model; M. Fleischer for help with programming the fluid finite-element model; R. G\"artner and A. Voigt for discussions of the finite-element model and stochastic computations; M. Lenz for discussions of stochastic computations and the analytic derivation of fluid-mediated interactions; and O. Ahmad, D. Andor, and M. O. Magnasco for discussions about data analysis. This research was funded by National Institutes of Health grant DC000241. Computational resources were provided by the Center for Information Services and High Performance Computing of the Technische Universit\"at Dresden. J.B. was supported by grants Gr~1388/14 and Vo~899/6 from the Deutsche Forschungsgemeinschaft. A.S.K. was supported by the Howard Hughes Medical Institute, of which A.J.H. is an Investigator. Fig.~\ref{fig:drag} is taken from reference~\cite{Kozlov2011} and Figs.~\ref{fig:ref}, \ref{fig:cyl}, and \ref{fig:fem} from the associated supplemental material with some modifications.
\end{theacknowledgments}

\bibliographystyle{moh2011}

\end{document}